\documentclass[twoside,twocolumn,9pt]{article}
\usepackage{extsizes}
\usepackage[super,sort&compress,comma]{natbib} 
\usepackage[version=3]{mhchem}
\usepackage[left=1.5cm, right=1.5cm, top=1.785cm, bottom=2.0cm]{geometry}
\usepackage{balance}
\usepackage{mathptmx}
\usepackage{sectsty}
\usepackage{graphicx} 
\usepackage{amssymb}
\usepackage{lastpage}
\usepackage[format=plain,justification=justified,singlelinecheck=false,font={stretch=1.125,small,sf},labelfont=bf,labelsep=space]{caption}
\usepackage{float}
\usepackage{fancyhdr}
\usepackage{fnpos}
\usepackage[english]{babel}
\addto{\captionsenglish}{%
  
}
\usepackage{array}
\usepackage{droidsans}
\usepackage{charter}
\usepackage[T1]{fontenc}
\usepackage[usenames,dvipsnames]{xcolor}
\usepackage{setspace}
\usepackage[compact]{titlesec}
\usepackage{hyperref}

\usepackage{epstopdf}

\definecolor{cream}{RGB}{222,217,201}

\usepackage{color}
\definecolor {myc} {rgb} {0,0,0}  

\definecolor {myc2} {rgb} {0,0,0} 
\begin{document}

\pagestyle{fancy}
\thispagestyle{plain}
\fancypagestyle{plain}{
\renewcommand{\headrulewidth}{0pt}
}

\makeFNbottom
\makeatletter
\renewcommand\LARGE{\@setfontsize\LARGE{15pt}{17}}
\renewcommand\Large{\@setfontsize\Large{12pt}{14}}
\renewcommand\large{\@setfontsize\large{10pt}{12}}
\renewcommand\footnotesize{\@setfontsize\footnotesize{7pt}{10}}
\makeatother

\renewcommand{\thefootnote}{\fnsymbol{footnote}}
\renewcommand\footnoterule{\vspace*{1pt}%
\color{cream}\hrule width 3.5in height 0.4pt \color{black}\vspace*{5pt}} 
\setcounter{secnumdepth}{5}

\makeatletter 
\renewcommand\@biblabel[1]{#1}            
\renewcommand\@makefntext[1]%
{\noindent\makebox[0pt][r]{\@thefnmark\,}#1}
\makeatother 
\renewcommand{\figurename}{\small{Fig.}~}
\sectionfont{\sffamily\Large}
\subsectionfont{\normalsize}
\subsubsectionfont{\bf}
\setstretch{1.125} 
\setlength{\skip\footins}{0.8cm}
\setlength{\footnotesep}{0.25cm}
\setlength{\jot}{10pt}
\titlespacing*{\section}{0pt}{4pt}{4pt}
\titlespacing*{\subsection}{0pt}{15pt}{1pt}

\fancyfoot{}
\fancyfoot[LO,RE]{\vspace{-7.1pt}\includegraphics[height=9pt]{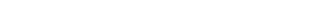}}
\fancyfoot[CO]{\vspace{-7.1pt}\hspace{13.2cm}\includegraphics{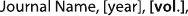}}
\fancyfoot[CE]{\vspace{-7.2pt}\hspace{-14.2cm}\includegraphics{head_foot/RF}}
\fancyfoot[RO]{\footnotesize{\sffamily{1--\pageref{LastPage} ~\textbar  \hspace{2pt}\thepage}}}
\fancyfoot[LE]{\footnotesize{\sffamily{\thepage~\textbar\hspace{3.45cm} 1--\pageref{LastPage}}}}
\fancyhead{}
\renewcommand{\headrulewidth}{0pt} 
\renewcommand{\footrulewidth}{0pt}
\setlength{\arrayrulewidth}{1pt}
\setlength{\columnsep}{6.5mm}
\setlength\bibsep{1pt}

\makeatletter 
\newlength{\figrulesep} 
\setlength{\figrulesep}{0.5\textfloatsep} 

\newcommand{\topfigrule}{\vspace*{-1pt}%
\noindent{\color{cream}\rule[-\figrulesep]{\columnwidth}{1.5pt}} }

\newcommand{\botfigrule}{\vspace*{-2pt}%
\noindent{\color{cream}\rule[\figrulesep]{\columnwidth}{1.5pt}} }

\newcommand{\dblfigrule}{\vspace*{-1pt}%
\noindent{\color{cream}\rule[-\figrulesep]{\textwidth}{1.5pt}} }

\makeatother

\twocolumn[
  \begin{@twocolumnfalse}
{\includegraphics[height=30pt]{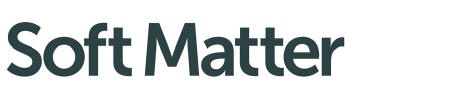}\hfill\raisebox{0pt}[0pt][0pt]{\includegraphics[height=55pt]{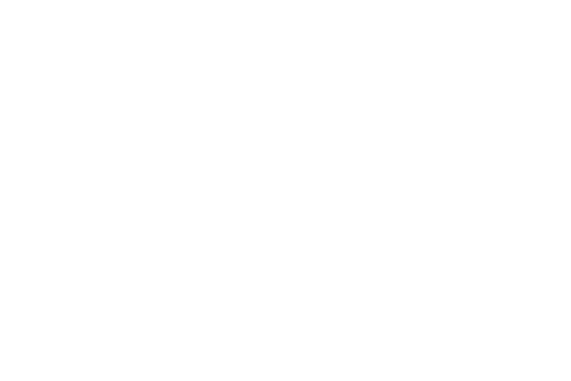}}\\[1ex]
\includegraphics[width=18.5cm]{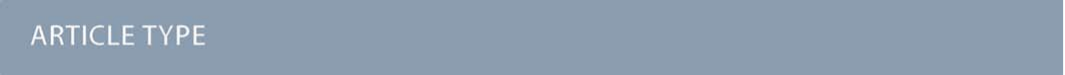}}\par
\vspace{1em}
\sffamily
\begin{tabular}{m{4.5cm} p{13.5cm} }

\includegraphics{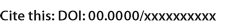} & \noindent\LARGE{\textbf{A double rigidity transition rules the fate of drying colloidal drops}} \\
\vspace{0.3cm} & \vspace{0.3cm} \\

 & \noindent\large{Matteo Milani\textit{$^{a}$}, Ty Phou\textit{$^{a}$},Christian Ligoure\textit{$^{a}$}, Luca Cipelletti\textit{$^{a}$} and Laurence Ramos\textit{$^{a}$}} \\

\includegraphics{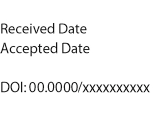} & \noindent\normalsize{The  evaporation of drops of colloidal suspensions plays an important role in numerous contexts, such as the production of powdered dairies, the synthesis of functional supraparticles, and virus and bacteria survival in aerosols or drops on surfaces. The presence of colloidal particles in the evaporating drop eventually leads to the formation of a dense shell that may undergo a shape instability. Previous works propose that, for drops evaporating very fast, the instability occurs when the particles form a rigid porous solid, constituted of permanently aggregated particles at random close packing. To date, however, no measurements could directly test this scenario and assess whether it also applies to drops drying at lower evaporation rates, severely limiting our understanding of this phenomenon and the possibility of harnessing it in applications. 
Here, we combine macroscopic imaging and space- and time-resolved measurements of the microscopic dynamics of colloidal nanoparticles in drying drops {\color{myc} sitting on a hydrophobic surface}, measuring the evolution of the thickness of the shell and the spatial distribution and mobility of the nanoparticles. We find that, above a threshold evaporation rate, the drop undergoes successively two distinct shape instabilities, {\color{myc}  invagination and cracking. While permanent aggregation of nanoparticles accompanies the second instability}, as hypothesized in previous works on fast-evaporating drops, we show that the first one results from a reversible glass transition of the shell, unreported so far. We rationalize our findings and discuss their implications in the framework of a unified state diagram for the drying of colloidal drops {\color{myc} sitting on a hydrophobic surface}.} \\

\end{tabular}

 \end{@twocolumnfalse} \vspace{0.6cm}

  ]

\renewcommand*\rmdefault{bch}\normalfont\upshape
\rmfamily
\section*{}
\vspace{-1cm}


\footnotetext{\textit{$^{a}$~Laboratoire Charles Coulomb (L2C), Universit\'e Montpellier, CNRS, Montpellier, France. E-mail: laurence.ramos@umontpellier.fr}}

\footnotetext{\dag~Electronic Supplementary Information (ESI) available: [details of any supplementary information available should be included here]. See DOI: 10.1039/cXsm00000x/}

\footnotetext{\ddag~Additional footnotes to the title and authors can be included \textit{e.g.}\ `Present address:' or `These authors contributed equally to this work' as above using the symbols: \ddag, \textsection, and \P. Please place the appropriate symbol next to the author's name and include a \texttt{\textbackslash footnotetext} entry in the the correct place in the list.}



\section{Introduction}

Drying drops of colloidal suspensions are an extremely intriguing phenomenon, since even a small amount of suspended particles can dramatically alter the fate of a drying drop: rather than shrinking continuously as drops of a simple fluid would do, colloidal drops typically undergo mechanical instabilities, such as buckling~\cite{tsapis2005onset, lintingre2015controlling,lintingre2016control}, air invagination~\cite{pauchard2004invagination}, and cracking~\cite{giorgiutti-dauphine_elapsed_2014,lyu2019final}. It is widely acknowledged that the fate of the drops is intimately linked to the formation of a dense colloidal shell. To rationalize whether a shell may form or not, one usually introduces a Péclet number, $Pe$ (see~\nameref{matmet}), which compares the efficiency of evaporation-induced advection to that of diffusion of the colloids. At low $Pe$, the colloids are essentially free to diffuse over the whole volume of the drop and thus are homogeneously distributed. By contrast, at high $Pe$ advection dominates over diffusion and the colloids accumulate at the liquid-air interface as the interface recedes, leading to the formation of a shell.

Processes such as inkjet printing or spray drying involve the evaporation of drops of complex fluids, at very high $Pe$. In these applications the drops dry in nearly boundary-free conditions, which can be reproduced in the lab by suspending a drop over a hot surface using the Leidenfrost effect~\cite{tsapis2005onset,lintingre2015controlling}. In these drying conditions, where $Pe \simeq 10^4$, experiments suggest that shape instability is due to the formation of a colloid-rich, rigid shell. It has been proposed that in the shell capillary forces push the colloidal particles close enough to overcome the repulsive interactions that normally maintain the suspension in a fully dispersed state~\cite{tsapis2005onset}, resulting in the irreversible particle aggregation, due to short-ranged van der Waals interactions. In this scenario, the shell is a rigid porous solid formed by permanently aggregated colloids, and the particle volume fraction in the shell is that of random close packing, e.g. $\phi_{\rm{shell}}=\phi_{\rm{rcp}}\approx0.64$ for spherical,  monodisperse particles. 

However, many drying processes relevant to biology or in industrial applications occur at $Pe$ orders of magnitude lower than those attained using the Leidenfrost effect. Examples include virus survival in aerosols~\cite{huynh2022evidence} $Pe \sim 10^{-4}-10^{-3}$, the production of powdered dairies~\cite{sadek2013shape} $Pe \sim 70$, or the preparation of functional supraparticles for catalysis, photonics, or sensing applications~\cite{liu2019tuning} $Pe \sim 0.1-1$,~\cite{kuncicky_surface-guided_2008} $Pe \sim 300$,~\cite{marin_building_2012} $Pe\sim 10-300$, ~\cite{sperling2014controlling} $Pe \sim 1-100$,~\cite{sekido2017controlling} $Pe \sim 10^{-3}-10^{-2}$. Even in these mild drying conditions, drops may show peculiar shape instabilities while drying, which have been mainly investigated in specific geometries (see~\cite{sadek_drying_2015} for a review), with the drops deposited on a surface (sessile drops)~\cite{basu_towards_2016,both2019role} or acoustically levitated~\cite{miglani2015sphere}, or hanging to a thin filament~\cite{fu_single_2012,sadek_drying_2015}. These works aimed at elucidating the effects on the morphology of the drying drop of several parameters, such as the suspension volume fraction and rheological properties,  the substrate type and, most importantly, the Péclet number. Unfortunately, no microscopic insight could be provided; consequently, these works had to rely on the untested assumption that the same physical scenario as the one described for high-$Pe$ drying also holds at lower $Pe$, namely that shape instability results from the irreversible aggregation of colloids at the drop periphery. To investigate the existence, evolution, and role of a rigid shell, several approaches have been adopted, including measuring the drop optical density~\cite{kuncicky_surface-guided_2008}, probing its structure with X-ray scattering using a synchrotron micro-focused beam~\cite{sen_slow_2007, sen2014probing, bahadur2015colloidal}, and imaging the drop {\textcolor{myc} {illuminated with a laser-sheet~\cite{bansal_universal_2015}}}, or with confocal~\cite{wooh_synthesis_2015,sperling2016understanding} or electron~\cite{tsapis2005onset,miglani2015sphere,bansal_universal_2015} microscopy, the latter being limited to the \textit{postmortem} analysis of fully dried drops. Unfortunately, these works could not provide detailed information on the shell thickness and its structural or dynamics properties, although they did confirm the existence of a shell. 

Experiments in a quasi-$2$D geometry allow for an easier characterization of the drying process. A few noticeable works provided quantitative measurements of the spatial distribution of the colloids during the evaporation of drops confined between two closely spaced plates. Using interferometry and a Raman confocal micro-spectrometer, the authors of Refs.~\cite{bouchaudy2019drying,sobac_collective_2020} measured the time evolution of the colloid distribution in the drop: no formation of a solid shell could be unambiguously detected and the colloid volume fraction was found to be smaller than $\phi_{\rm{rcp}}$ at all times. By contrast, Boulogne \textit{et al.} could detect the formation of a shell and followed the  evolution of its thickness~\cite{boulogne_buckling_2013}, but unfortunately the concentration of colloids in the shell could not be determined. Overall, although experiments in $2$D are insightful {\color{myc2} {and provide a consistent picture, a direct connection with drying in a $3$D geometry remains difficult to be established. }}

Here, we go beyond previous studies by investigating the evaporation of 3D colloidal drops {\color{myc} {sitting on a hydrophobic surface}}, at intermediate Péclet numbers, with a unique combination of macroscopic imaging and space- and time-resolved measurements of the microscopic dynamics of the colloidal suspension within the drop. We find that, quite generally, drying drops undergo two distinct, successive shape instabilities, a result unreported so far. The second instability, {\color{myc} {a cracking instability, with air penetration in the material is concomitant with the irreversible aggregation of colloidal particles, consistently}} with previous investigations at high $Pe$. The first instability occurs at a stage where the drying process is still reversible and the colloids can be fully re-dispersed. By measuring the microscopic dynamics in the shell, we demonstrate that this instability is triggered by a colloidal glass transition. Finally, we propose a state diagram that rationalizes our experiments and previous measurements, providing a general framework with implications ranging from the synthesis of supraparticles to the survival of biological entities in drying drops.

\section{Materials and Methods}
\label{matmet}
\subsection{Setup}

\begin{figure}
\includegraphics[width=1\linewidth]{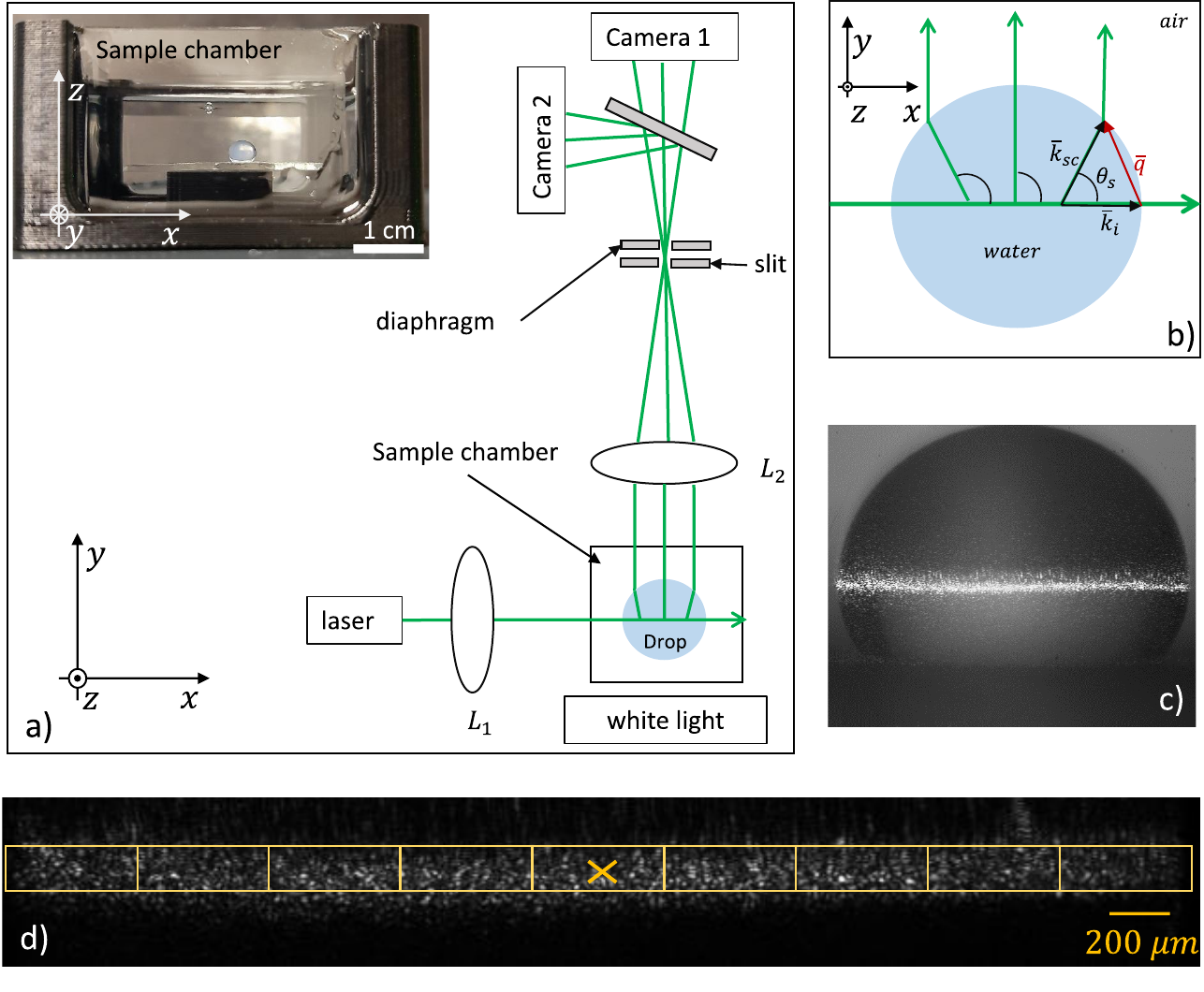}
\caption{\label{fig:set-up} (a) Scheme of the dynamic light scattering setup used to measure the dynamics of colloids in a drop. The incoming laser beam propagates along the $x$ axis and is mildly focused by the lens $L_1$ in the middle of the sample. The lens $L_2$ forms an image of the scattering volume on the sensor of Camera 1. In addition to the laser beam, white light provided by a LED screen illuminates the sample, in order to form on Camera 2 an image of the full drop, to monitor its macroscopic evolution. By combining a vertical slit, parallel to the $z$ direction, and a diaphragm, we ensure that both cameras are illuminated by light exiting the drop along the $y$ direction, to within acceptance angles of $0.55$ deg and  $1.65$ deg in the plane $(x,y)$ and $(y,z)$, respectively. The sample is placed in a cubic glass box of side $4$ cm, within which controlled humidity conditions are imposed. The inset shows an image of the sample chamber containing a drop (diameter $3.5$ mm) of a colloidal suspension sitting on a hydrophobic surface and surrounded by oil, used to stop evaporation. (b) Schematic top view of the relevant optical rays for a drop surrounded by air. The scattered light is refracted at the drop-air interface such that a somehow different scattering angle $\theta_s$ corresponds to each position $x$. (c) Image formed on Camera 2 of a drop of colloidal suspension illuminated simultaneously by the laser beam and by white light. (d) Cropped speckle image ($100 \times1300$ pixels$^2$), formed on Camera 1 by the light scattered in the drop and collected by the lens $L_2$. The yellow rectangles indicate the Regions of Interest for the space-resolved data analysis and the cross corresponds to the drop center ($x=0$).}
\end{figure}

We show in Fig.~\ref{fig:set-up}a a scheme of the Photon Correlation Imaging (PCI~\cite{duri2009resolving}) light scattering apparatus specifically designed to probe, with both spatial and temporal resolutions, the dynamics of colloids confined in a spherical drop of radius $R$ of the order of $1$ mm (see~\cite{SM} for details). A laser beam (\textit{in-vacuo} wavelength $\lambda =532$ nm, maximum power of $2$ W) propagates along the $x$ axis and is mildly focused in the drop center by the lens $L_1$, such that the scattering volume is approximately a thin cylinder of radius $w = 60~\mu$m oriented  along a drop diameter, with the beam waist $w << R$. An image of the scattering volume is formed on the detector of a CMOS camera, using the lens $L_2$. A vertical slit and a diaphragm are placed in the focal plane of $L_2$, such that the image is formed only by light exiting the sample chamber within a thin solid angle centered around the $y$ direction~\cite{SM}. The setup combines features of imaging and light scattering: each location on the detector corresponds to a well-defined position in the sample, as in conventional imaging, but, contrary to conventional imaging, the image formed on the detector result only from light scattered at a chosen angle and thus at a specific scattering vector $q$.

As schematically shown in Fig.~\ref{fig:set-up}b, the light propagating towards the detector is refracted at the drop-air interface. This has two consequences: first, the position along the $x$ axis of a ray exiting the drop is in general different from the $x$ coordinate of the sample volume from which that scattered light originates. Throughout the paper, refraction effects are corrected: the coordinate $x$ refers to the actual position within the scattering volume, $x=0$ being the drop center. Second, the scattering angle $\theta_S$ associated to optical rays exiting the drop along the $y$ direction depends on $x$. Using Snell's law, we calculate the $x$-dependence of $\theta_S$ and of the modulus $q$ of the scattering vector, with $q = 4\pi n/\lambda \sin(\theta/2)$, $n=1.33$ being the refractive index of the solvent (water). We find that the $x$-dependence of $q$ is mild: typically, $q$ ranges from $18$ $\mu$m$^{-1}$ to $26$ $\mu$m$^{-1}$ depending on the position $x$ within the drop, see~\cite{SM} for a detailed discussion.

\subsection{Data Analysis}
\label{subsec:data_analysis}

To quantify the dynamics, we use Camera 1 to take a time series of images of the scattered light. The images have a speckled appearance, as shown in Fig.~\ref{fig:set-up}d. Camera 1 collects images of size $(512 \times 1088)$ pixels$^2$, allowing for the acquisition of videos at rates up to 200 Hz, using the acquisition scheme of Ref.~\cite{philippe2016efficient}. Each image is divided into Regions of Interest (ROIs), typically rectangles of dimension  $(30 \times100)$ pixels$^2$, corresponding to about $(70 \times 260)$ $\mu$m$^2$ in the sample. The size of the ROIs is chosen so that each ROI contains around $100$ speckles, which provides an acceptable statistical noise~\cite{duri_timeresolvedcorrelation_2005}. The local dynamics within a given ROI are quantified by a two-time degree of correlation~\cite{duri_timeresolvedcorrelation_2005}:
\begin{equation}
    c_{I}(t,\tau,x) = A \frac{\langle I_p(t)I_p(t+\tau) \rangle_{x} }{\langle I_p(t)\rangle_{x}\langle I_p(t+\tau) \rangle_{x}}-1
\end{equation}
with $\tau$ the time delay between images, $I_p(t)$ the intensity of the $p$-th pixel of the ROI at time $t$, and $\langle ...\rangle_{x}$ an average over all pixels of an ROI centred around $x$. The prefactor $A$ is chosen such that the intensity correlation function $g_2(\tau, x) - 1$, obtained by averaging $c_{I}(t,\tau,x)$ over a suitable time interval, tends to $1$ for $\tau\to 0$. For a dilute suspension of Brownian particles, $g_2(\tau)-1$ is a simple exponential decay: $g_2(\tau)-1 = e^{-\tau/\tau_D}$. For systems with more complex dynamics, $g_2(\tau)-1$ deviates from this simple shape. In the general case, the relaxation time of $g_2-1$ can be conveniently obtained experimentally from $\tau_D = \int_{0}^{\tau_{max}}\left[g_2(\tau)-1\right] \textrm{d}\tau$, where $\tau_{max}$ is the shortest delay for which $g_2(\tau)-1$ decays to zero to within experimental uncertainties.

\subsection{Samples}

We use commercial silica nanoparticles (Ludox TM-50 from Sigma Aldrich) suspended in water. {\textcolor{myc} {The batch is used as received}} The nanoparticle diameter is $2a=25$ nm as measured by dynamic light scattering (DLS). For all experiments, the initial volume fraction of NPs in the drop is $\phi=0.31$.  The suspension is transparent to visible light. Hence, when passing through the drop, the laser light is scattered at most once, thus avoiding multiple scattering. Working in the single scattering regime is essential for measuring the dynamics with space-resolution. The NPs are charge stabilized and thus interact through a soft repulsive Yukawa potential. The Debye length is of the order of $20$ nm~\cite{philippe2018glass}, comparable to the NP radius. {\color{myc} {The dynamics of the same nanoparticles have been previously measured at the same $q$ vector for bulk quiescent suspensions \cite{philippe2018glass}. In the supercooled regime, from $\phi=0.32$ to the glass transition $\phi=0.39$, the relaxation time of the system grows steeply with volume fraction (see also Fig.~{\color{myc} {11}} in~\cite{SM}).}}

\subsection{Experimental protocol}

A drop of volume $30$ $\mu$L and radius $R_0=1.9$ mm is gently deposited on a hydrophobic surface, forming a contact angle of $120$ deg. The drop is surrounded by air or by silicon oil (47 V 100, VWR chemicals, refractive index $=1.403$). Temperature is controlled at $(22\pm1)^{\circ}$C. To control the relative humidity RH, we fill the bottom of the sample chamber (a cubic glass box of side $4$ cm) with different chemicals or saturated salt solutions. We vary RH in the range $20$\% to $94$\%, thereby imposing different evaporation rates to the drop. {\textcolor{myc} {The drop evaporates in an intermediate mode between the constant contact radius mode and the constant contact angle mode.}} The characteristic evaporation time, $\tau_{\rm{ev}}$, is defined as the time to fully evaporate the drop, should it not contain nanoparticles. As in previous works~\cite{langmuir1918evaporation, sobac2015comprehensive}, $\tau_{\rm{ev}}$ is evaluated by extrapolating the  small-$t$ behaviour of $R(t)^2$, with $R(t)$ the time-dependent drop radius and $t=0$ the time at which evaporation starts (see Fig.~4 in~\cite{SM}).

In our experiments,  $\tau_{\rm{ev}}$ varies from $(465\pm10)~\textrm{s}$ to $(43960\pm20)~\textrm{s}$  (see~\cite{SM} for details). The relevant non-dimensional Péclet number, $Pe=\tau_{\rm{mix}}/\tau_{\rm{ev}}$, compares the evaporation time to $\tau_{\rm{mix}}=R_0^2/D$, the characteristic mixing time due to Brownian diffusion over a distance equal to the drop initial size. {\color{myc2} {We take $D = 4.31 \times 10^{-12}~\mathrm{m^2s^{-1}}$, the diffusion coefficient of the NPs for $\phi=0.31$~\cite{philippe2018glass} }}. In our experiments, $Pe$ varies from {\color{myc2} {$63$ to $5754$}}. The reduced time  $\tilde{t}$ is defined as $\tilde{t}=t/\tau_{\rm{ev}}$.

\section{Results}

\subsection{Shape Instabilities and Nanoparticles  Aggregation}\label{Shapeinstabilityandaggregation} 
\begin{figure}
\includegraphics[width=1\linewidth]{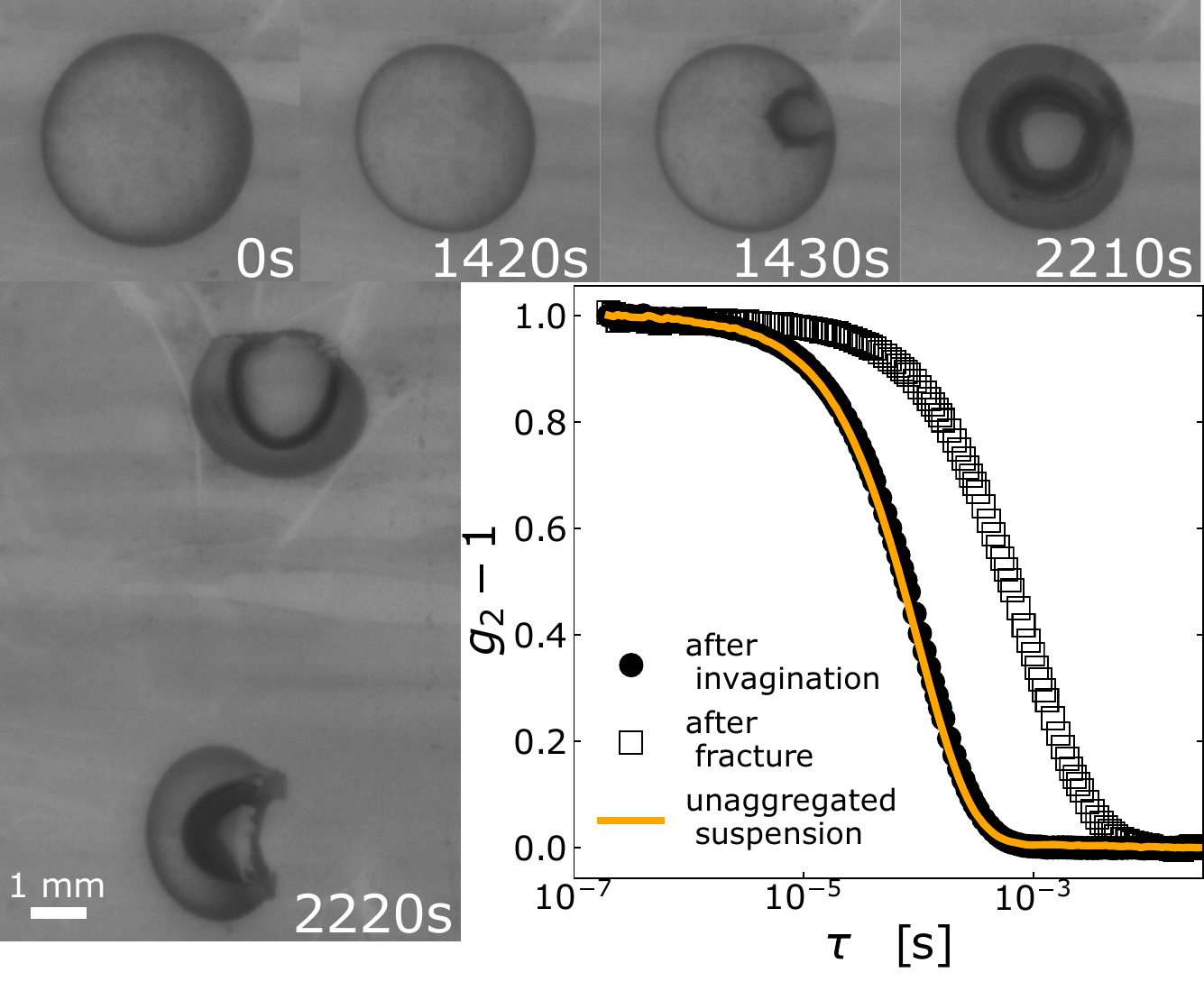}
\caption{\label{fig:imaging_top} \textbf{Shape instabilities and colloidal aggregation.} Images: sequence of top views of a drop of colloidal suspension drying at $Pe=399$. {\textcolor{myc} {A movie is provided in~\cite{SM}.}}The scale is the same in all images and the scale bar represents $1$ mm. Images are labelled by the time $t$ (in s) since depositing the drop on an hydrophobic substrate. The onset of the shape instability is visible at $t=1430$ s;  subsequently the invaginated region grows until the drop suddenly {\color{myc} {cracks}} at $t=2220$ s. Graph: intensity correlation functions measured by conventional DLS probe the aggregation state of the colloids. Line: batch suspension. Filled circles: drop collected in the invaginated state ($t= 2210$ s) and redissolved in water. Empty circles: fractured drop ($t=2220$ s) immersed in water and sonicated. All DLS curves are collected at a NPs volume fraction $\phi=0.014$.}
\end{figure}

To study the macroscopic behavior of drying drops we deposit $30$ $\mu$L of a silica nanoparticles suspension (NPs, see section~\nameref{matmet} for details) on a hydrophobic surface.  The surface and the drop are in a closed chamber with controlled relative humidity. Figure~\ref{fig:imaging_top} displays a representative sequence of top views of a drop that dries at $Pe=399$ (see section~\nameref{matmet} for details on the determination of $Pe$). Right after deposition and up to $t \approx 1420$ s 
the drop maintains {\textcolor{myc2} {a spherical cap shape}} while shrinking. At $t=1430$ s, a shape instability in the form of air invagination is clearly seen. The shape instability develops with time, see $t=2210$ s in Fig.~\ref{fig:imaging_top}, until a second shape transition occurs with the sudden macroscopic fracture of the drop, $t=2220$ s in Fig.~\ref{fig:imaging_top}. By running experiments at $Pe$ in the range $63-5754$, we establish that a similar two-state transition always occurs for $Pe>70$, whereas drops dried at lower rate ($Pe=63, 70$) do not exhibit the first shape instability but only a macroscopic fracture.

As mentioned above, for drops rapidly dried using the Leidenfrost effect the onset of the shape instability results from colloids aggregating to form a porous solid shell~\cite{tsapis2005onset}. 
In our experiments conducted at much lower drying rates, drops that underwent shape instability but did not {\color{myc} {crack}} yet (e.g. at $t= 2210$ s in Fig.~\ref{fig:imaging_top}) can be fully re-dispersed in water, as checked by visual inspection, while 
once the drop has macroscopically fractured ($t= 2220$ s in Fig.~\ref{fig:imaging_top}), the fragments do not dissolve in water even after 48 hours. 
We confirm these macroscopic observations using conventional dynamic light scattering (DLS~\cite{pecora2000dynamic}) to measure the size of the particles or aggregates after re-dispersing the drop in water. For the invaginated sample, we find that the intensity correlation function $g_2-1$ measured by DLS is indistinguishable from that of a freshly prepared suspension of unaggregated NPs. Due to strong Van der Waals attractions, aggregated NPs would not re-disperse spontaneously: we thus conclude that no particle aggregation occurs throughout the development of the shape instability. By contrast, the correlation function for the fractured sample decays on a time scale one order of magnitude larger, even after sonicating the fragments for one hour. In DLS-based particle sizing, the decay time of $g_2-1$ is proportional to the particle size; thus, these data demonstrate that the NPs irreversibly aggregate just before the drop fractures, forming strong bonds that are only partially broken by vigorous sonication.

These experiments show that nanoparticles eventually do aggregate during the drying of drops of colloidal suspensions, even for $Pe$ lower than the ones studied in the seminal work of Tsapis and Weitz~\cite{tsapis2005onset}. However, in the {\textcolor{myc}moderate to high  {$70 < Pe \le\sim6000$}} range, the shape instability sets in before the NPs undergo aggregation, challenging the commonly proposed scenario that attributes quite generally shape instability to colloidal aggregation, including at low $Pe$~\cite{basu_towards_2016,sobac2019mathematical,boulogne_buckling_2013}. Our findings thus raise the question of the physical mechanism responsible for the onset of a shape instability for slowly evaporating drops. Previous works suggested that the formation of a shell plays a key role. In the following, we use a custom  light scattering setup to measure \textit{in-situ} the shell formation and growth, probing the microscopic dynamics of the NPs within the drying drop. The setup is based on Photon Correlation Imaging (PCI ~\cite{duri2009resolving}), a space- and time-resolved variant of DLS, see section~\nameref{matmet} and \cite{SM} for more details.

\subsection{Evolution of the Shell Thickness}\label{Shell thickness and its evolution}

\begin{figure*}
\includegraphics[width=1\linewidth]{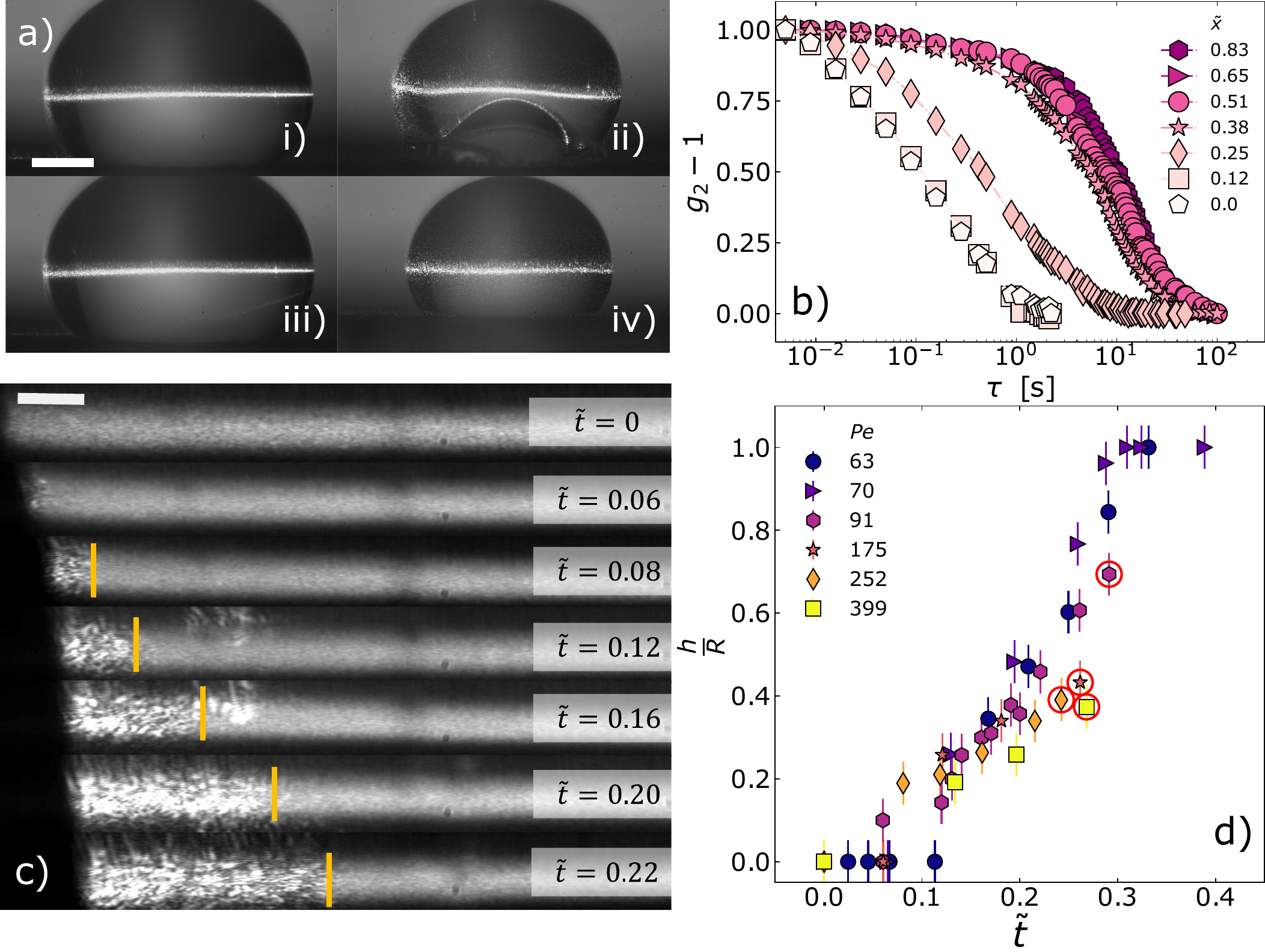}
\caption{\label{fig:shell} \textbf{Shape Instability and Shell Formation} (a) Side views of drops drying at $Pe=252$ (i, ii) and $Pe=70$ (iii, iv), respectively. The drops are illuminated by both white light and a thin laser beam; the images are taken by Camera 2 as shown in Fig.~\ref{fig:set-up}. Images i) and iii) are taken at time $t=0$; ii) and iv) at reduced times $\tilde{t}=0.242$ and $\tilde{t}=0.3$, respectively. In ii), nanoparticles accumulating at the drop-air interface created by the shape instability are visible because they are illuminated by light scattered by the volume directly illuminated by laser beam. Note that no instability occurs for the drop drying at $Pe=70$ (iii, iv). The scale bar represents 1 mm. (b) Intensity correlation functions taken simultaneously at $\tilde{t}=0.23$ at different positions, from the drop center ($\tilde{x}=0$) to near the edge, located at $\tilde{x}=1$, for a drop drying at $Pe=91$. (c) Formation and evolution of the shell during drop evaporation for a sample drying at $Pe = 91$. Speckle images averaged over $\Delta t = 4$ s and taken at different reduced times, as indicated by the labels. The vertical yellow lines separate the homogeneous, blurred region corresponding to the core of the drop with fast dynamics, to the shell region where the speckles are still visible after averaging, indicating slow dynamics. For clarity, only the left half of the drop is shown; the center of the drop sits on the right edge of the images.  
The scale bar represents $0.1$ mm. (d) Shell thickness normalized by the drop radius $R(t)$, as a function of the reduced time $\tilde{t}$. Different colors and symbols correspond to different Péclet numbers in the range $63-399$, as indicated by the labels. {\color{myc} {Red circles represent the instant of the invagination, which occurs at $\tilde{t}=0.27, 0.24, 0.26,0.29$ for $Pe=399,252,175,91$ respectively}}. Error bars results from the uncertainty in the determination of the drop radius and of the shell thickness in the speckle images.}
\end{figure*}

Figure~\ref{fig:shell}a displays side views of drops that are simultaneously illuminated by white light, for conventional imaging, and by a thin laser beam, for PCI. ({\textcolor{myc} {A Side view movie for a drop illuminated only by white light is provided in~\cite{SM}}}.) Consistently with the top views of Fig.~\ref{fig:imaging_top}, a shape instability is observed for the experiment run at $Pe=252$, whereas no instability takes place on the time scale of the experiment for a drop drying at $Pe=70$.  As detailed in section~\nameref{matmet}, we compute the $g_2 - 1$ intensity correlation functions from the time fluctuations of the speckle pattern generated by the interference of light scattered by the NPs. Typical $g_2 - 1$ functions are shown in Fig.~\ref{fig:shell}b,
for a sample that dries at $Pe = 91$ and for a fixed reduced time $\tilde{t}=0.23$, defined as $\tilde{t}=t/\tau_{\rm{ev}}$, with $t$ the actual time since the deposition of the drop on the substrate and $\tau_{\rm{ev}}$ the evaporation time (see section~\nameref{matmet} and Supporting Information~\cite{SM} for details). The various curves are measured simultaneously for different positions along the drop diameter, identified by the reduced coordinate $\tilde{x}=x/R$, with $x$ the position in the drop ($x=0$ corresponds to the drop center) and $R$ the drop radius. The decay time of $g_2-1$ corresponds to the time it takes to relax particle density fluctuations of wavelength $1/q \approx (40-55)$ nm via the microscopic motion of the NPs ($q$ is the magnitude of the scattering vector, see section~\nameref{matmet}). 

The intensity correlation functions strongly depend on $\tilde{x}$, with a faster decay in the center of the drop (time scale of the order of $0.1$ s for $\tilde{x}= 0$) and a much slower one near the edge (time scale of tens of seconds for $\tilde{x} = 0.83$). Quite generally, the dynamics of a colloidal suspension strongly slow down as the particle concentration increases~\cite{pusey_observation_1987}. Therefore, the spatial variation of the NPs mobility suggests the formation of a colloid-rich shell that surrounds a diluted core. {\color{myc} {In the whole paper, we define as shell, the localized region close at the drop periphery where the NP dynamics is at least two orders of magnitude slower than that in the core of the drop.}}

To quantify the shell thickness $h$, we average movies of the speckle pattern over a suitable time interval, $\Delta t =4$ s, intermediate between the decay time of $g_2-1$ in the core and that in the shell. Figure~\ref{fig:shell}c shows the results of such processing for a drop drying at an intermediate evaporation rate, $Pe = 91$. The fast-fluctuating core appears as a gray region, while speckles in the shell are clearly visible, because they are essentially frozen on the time scale  $\Delta t $ of the time average. Figure~\ref{fig:shell}c shows that the shell develops with time, thickening progressively. By analyzing the spatial variance of the intensity of the time-averaged images (details in the Supporting Information~\cite{SM}), we determine the boundary between the shell and the core, yellow lines in Fig.~\ref{fig:shell}c, checking that the value of $h$ thus obtained does not depend significantly on the choice of $\Delta t$, see~\cite{SM}. Here and in the following we show only the results for one half of the drop, for symmetry reasons.

Figure~\ref{fig:shell}d shows the evolution of $h$ normalized by the instantaneous radius of the drop, $R(t)$, as a function of the reduced time $\tilde{t}$. Overall, our findings are in qualitative agreement with the $Pe$ dependence of a shell, as determined through numerical calculations for {\textcolor{myc2} {3D spherical drops.}}~\cite{sobac2019mathematical}. For all investigated $Pe$, we find that the thickness increases with time, at a rate that slightly decreases as $Pe$ increases. The same data displayed in the Supporting Information~\cite{SM} as plots of $h/R$ versus absolute time $t$ show that the growth rate of the shell in absolute units is approximately the same for all samples, and that shell formation starts after a latency time that decreases with $Pe$. For the lowest $Pe$ investigated, $Pe\leq70$, $h/R$ reaches $1$, indicating that the shell invades the whole drop volume, and thus the spatial distribution of the nanoparticles becomes homogeneous at long times, {\textcolor{myc} {in agreement with experimental findings for 2D drops~\cite{sobac_collective_2020}.}} 
After reaching such a homogeneous state, the drop shrinks smoothly (see iii)-iv) in Fig.~\ref{fig:shell}a for a sample drying at $Pe=70$) until it abruptly fractures. By contrast, for drops that evaporate at a faster rate homogeneous shrinking is always stopped by the onset of a shape instability at the bottom of the drop (see i)-ii) in Fig.~\ref{fig:shell}a for a sample drying at $Pe=252$).  Interestingly, as indicated by red circles around the data points in Fig.~\ref{fig:shell}b, for $Pe\ge 175$ the onset of the shape instability occurs approximately at the same $\tilde{t}$ and $h/R$.

\subsection{Space-dependent Microscopic Dynamics and Nanoparticles Concentration}\label{Spatio-temporal dependence of the nanoparticles dynamics}

\begin{figure*}[ht]
\includegraphics[width=\textwidth,height=\textheight,keepaspectratio]{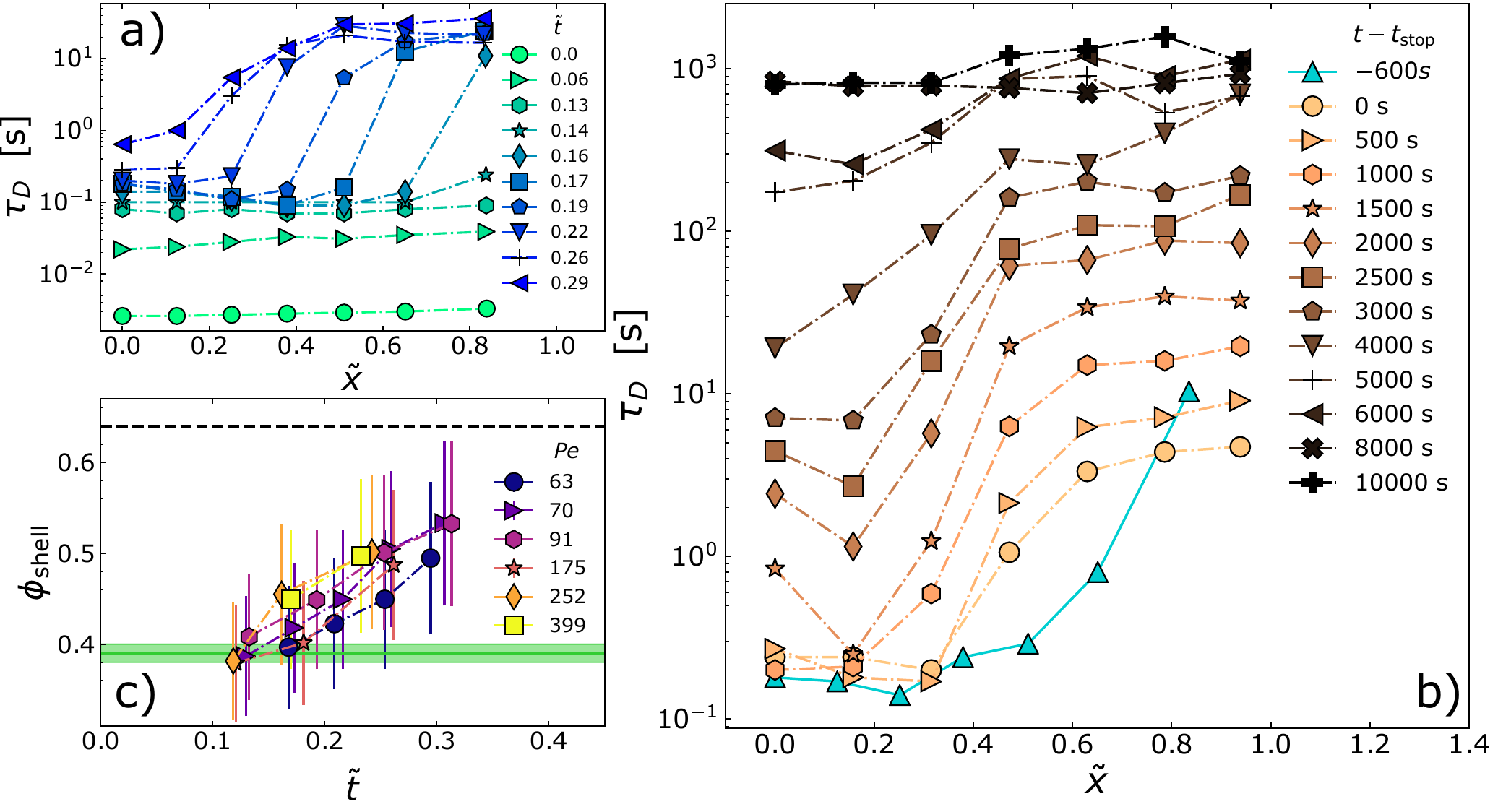}
\caption{\label{fig:dynamics} \textbf{Microscopic dynamics during drop evaporation} (a) Microscopic relaxation time as a function of the normalized position along the drop radius $\tilde{x}=x/R$, for $Pe=91$. Data are labelled by the reduced time $\tilde{t}$.  (b) Decay time of $g_2(\tau)-1$ as a function of position along the radius for a drop immersed in oil. The drop was previously dried at $Pe=91$ for $t_{\rm{stop}}=5880~\textrm{s}$ ($\tilde{t}=0.16$), leading to a decrease of its radius by $9$\%, before being immersed in oil to stop evaporation. Curves are labelled by the time since stopping evaporation. The up triangle symbols and solid line correspond to $\tau_D$ values $600$ s before the drop was immersed in oil.  (c) Volume fraction of the shell as a function of reduced time, as obtained from mass conservation and measurements of the shell thickness and core volume fraction. Symbols are data points for drops dried at different Péclet numbers, as indicated by the labels. The error bars are the uncertainty due to approximations on the drop and core shape, which have been considered spherical, see~\cite{SM}. Horizontal lines: volume fraction of the glass transition for the NPs used here (solid line, $\phi_g=0.39${\color{myc} {$\pm0.01$}}~\cite{philippe2018glass}, the green band shows uncertainties) and the random close packing volume fraction of monodisperse particles (dotted line, $\phi=\phi_{rcp}=0.64$), a lower bound for the actual volume fraction at rcp for our polydisperse nanoparticles~\cite{baranau_random-close_2014,trzaskus2016understanding}.}
\end{figure*}

To gain a deeper insight into the NPs microscopic dynamics and their spatial distribution, we calculate $\tau_D$, the space- and time-dependent time scale of the microscopic dynamics, obtained from the decay time of the intensity correlation functions $g_2-1$ following the procedure detailed in \nameref{matmet}. Figure~\ref{fig:dynamics}a displays $\tau_D$ for a representative experiment at $Pe= 91$, as a function of reduced position $\tilde{x}$ and for various reduced times $\tilde{t}$. At early times, the dynamics are spatially homogeneous and slow down with $\tilde{t}$ due to the overall increase of the volume fraction of the nanoparticles as the drop shrinks. At later times, by contrast, the dynamics at the edge of the drop become almost two orders of magnitude slower than those in the center, signaling the formation of a dense shell. Consistently with what seen in the time-averaged speckle images (Fig.~\ref{fig:shell}c), the region with slow dynamics expands with time, until the occurrence of a drop instability at $\tilde{t} = 0.30$. Results for drops drying at $Pe = 63, 70, 175, 252,~\mathrm{and}~399$ are shown in Fig.~{\color{myc} {10}} in~\cite{SM}. At $Pe=63, 70$ the spatial dependence of the dynamics is initially similar to that shown in Fig.~\ref{fig:dynamics}a for $Pe=91$. However, the drops do not display any instability and the nanoparticles eventually recover a homogeneous distribution within the drop, as demonstrated by the weak $\tilde{x}$-dependence of the dynamics at late times, $\tilde{t}\approx 0.3$. The drops then further dry, until abruptly cracking, at $\tilde{t}\approx0.39$ for $Pe=70$. By contrast, the samples evaporating more rapidly ($Pe=175, 252, 399$) form a shell since the early steps of evaporation, $\tilde{t}\approx0.08$. In these cases, the shell is thinner than that measured at lower $Pe$, but the contrast between the dynamics of the core and that of the shell, two orders of magnitude in $\tau_D$, is similar for all fast-evaporating samples.

Nanoparticle suspensions in a drying drop are not at thermodynamic equilibrium: we thus expect their dynamics to depend on both the (local) volume fraction, $\phi$, and the drying process. To disentangle the effect of $\phi$ from that of drying, we run an experiment where a drop first evaporates at $Pe = 91$, and then is immersed in oil to stop water evaporation, at $t = t_{\rm{stop}} = 5880~\mathrm{s}$ ($\tilde{t}_{\rm{stop}}=0.16$), before the occurrence of any instability. Figure~\ref{fig:dynamics}b shows the spatial dependence of the microscopic relaxation time $\tau_D$ just before stopping evaporation ($t=t_{\rm{stop}}-600~\textrm{s}$, up triangles and solid line) and for several times $t \geq t_{\rm{stop}}$. Remarkably, the relaxation time in the center of the drop, $\tau_D^{\rm{core}} \equiv \tau_D(\tilde{x}=0)$, remains essentially the same as the one during evaporation for at least $500~\textrm{s}$ after $t_{\rm{stop}}$. This indicates that evaporation did not impact significantly the dynamics in the center of the drop. Subsequently, $\tau_D^{\rm{core}}$ progressively increases, because of the re-dispersion of the nanoparticles from the shell to the core, which increases $\phi$ in the core. Consistently, we find that the dynamics become spatially homogeneous, provided that the drop is let equilibrating under no evaporation conditions for a long enough time. Indeed, for $t> t_{\rm{stop}}+8000~\textrm{s}$, we measure a constant $\tau_D=(949\pm 17)~\textrm{s}$ throughout the drop. ($8000$ s corresponds roughly to a $1.4$-fold the time the drop has evaporated before being immersed in oil). Using the $\phi$-dependence of the relaxation time measured for the same sample in quiescent, equilibrium conditions as a calibration curve (Ref.~\cite{philippe2018glass} and Fig.~{\color{myc} {11}} in~\cite{SM}), we find that this value of $\tau_D$ corresponds to a spatially homogeneous NPs volume fraction $\phi=0.385$, in excellent agreement with $\phi=0.386$ as estimated from mass conservation using the initial nanoparticle concentration and the measured drop volume reduction.

The experiment where drying was stopped proves that the relaxation time in the core is a robust proxy for the nanoparticle concentration and that $\phi_{\rm{core}}$ can be reliably calculated from reference data for quiescent suspensions~\cite{philippe2018glass,SM}. We further confirm this approach by measuring $\phi_{\rm{core}}$ at $t=0$, finding $\phi_{\rm{core}}=0.3014 \pm 0.008$ averaged over all experiments, in excellent agreement with $\phi = 0.312$, the nominal value of the volume fraction of the batch suspension used to prepare the drops.

Knowing the volume fraction in the core and the thickness of the shell (Fig.~\ref{fig:shell}b), we calculate the time-dependent volume fraction of the shell, $\phi_{\rm{shell}}$, using mass conservation and assuming a step-wise radial concentration profile, with constant $\phi$ in the core and the shell, respectively. Figure~\ref{fig:dynamics}c displays the time evolution of $\phi_{\rm{shell}}$ for all the Péclet numbers where a shell was observed. We find that $\phi_{\rm{shell}}$ steadily increases with time, in a mildly $Pe$-dependent manner. Surprisingly, we find that the shell is not a very dense medium composed of colloids at random close packing, but rather a concentrated suspension whose volume fraction depends only weakly on the rate at which the drop shrinks, and reaches at most $0.55$ when the drop undergoes the shape instability. This value is lower than the reference value for randomly closed packed monodisperse spheres ($\phi_{rcp} \approx 0.64$),  a lower bound for the actual volume fraction at rcp for our polydisperse nanoparticles~\cite{baranau_random-close_2014,trzaskus2016understanding}.

\subsection{Glassy Dynamics of the Shell}
\label{Glassydynamics}

\begin{figure}
\includegraphics[width=1\linewidth]{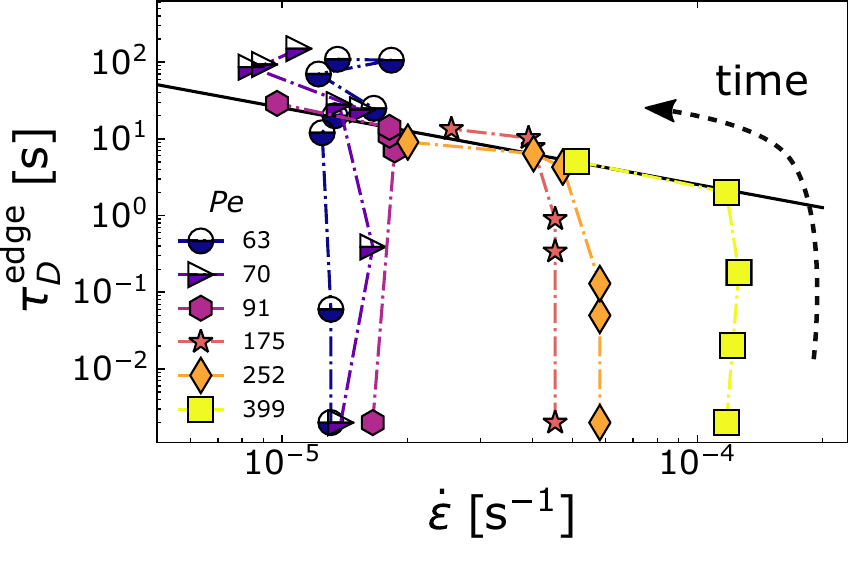}
\caption{\label{fig:taudvsstarinrate} \textbf{Shell dynamics driven by macroscopic shrinkage}. Microscopic relaxation time, $\tau_D^{\rm{edge}}$, of the nanoparticles located at the edge of the drop, as a function of the (time-dependent) macroscopic strain rate, $\dot{\epsilon}$, for various Péclet numbers as indicated in the legend. For all experiments, $\tau_D^{\rm{edge}}$ increases monotonically with time, as shown by the arrow for the sample at $Pe = 399$. Solid (resp., semi-filled) symbols refer to drops that eventually do (resp., do not) exhibit a shape instability. All data shown here are collected before any instability occurs.  The black solid line is a fit of $\tau_D^{\rm{edge}}=\Lambda\dot{\epsilon}^{-1}$ to the data for $Pe \ge 91$ in the final stages of evaporation, after the formation of a glassy shell, yielding $\Lambda = (2.5 \pm 0.1)\times 10^{-4}$.}
\end{figure}

The values of $\phi_{\rm{shell}}$ shown in Fig.~\ref{fig:dynamics}c confirm the \textit{post mortem} observations of Fig.~\ref{fig:imaging_top}, ruling out the hypothesis that the shape instability is due to the irreversible aggregation of the NPs in the shell. Values of $\phi_{\rm{shell}}$ in the range $0.5-0.6$, as measured just before the shape instability, are rather suggestive of a colloidal glass transition, upon which the dynamics of a colloidal suspension slow down dramatically and its mechanical behavior approaches that of a solid~\cite{pusey_observation_1987,philippe2018glass}, while the sample maintains an unaggregated structure, similar to that of a dense fluid. Indeed, we consistently find that $\phi_{\rm{shell}} \ge \phi_g=0.39${\color{myc} {$\pm0.01$}}, the glass transition volume fraction measured for the same system under quiescent conditions~\cite{philippe2018glass} (solid horizontal line in Fig.~\ref{fig:dynamics}c).

To gain insight into the origin of the shape instability, we inspect $\tau_D^{\rm{edge}}$, the microscopic relaxation time at the drop periphery. 
Surprisingly, Figure~\ref{fig:dynamics}a shows that $\tau_D^{\rm{edge}}$ never exceeds a few tens of s, orders of magnitude less than what expected for quiescent glassy suspensions at volume fractions comparable to those estimated for $\phi_{\rm{shell}}$~\cite{philippe2018glass,SM}. This apparent conundrum is solved by noticing that the dramatic acceleration of the dynamics is due to the strain field imposed by drying: Figure~\ref{fig:dynamics}b shows that stopping evaporation leads to an increase of $\tau_D^{\rm{edge}}$ by at least two orders of magnitude, compare the data at $\tilde{x} \lesssim 1$ just before $t_{\rm{stop}}$ (up triangles) and at $t=t_{\rm{stop}}+10^4~\textrm{s}$ (plus symbols).

We rationalize the relationship between microscopic dynamics and macroscopic shrinkage by displaying in Fig.~\ref{fig:taudvsstarinrate} $\tau_D^{\rm{edge}}$ as a function of the instantaneous drop strain rate $\dot{\epsilon}= \frac{1}{R} \left | \frac{dR}{dt} \right|$, for various Péclet numbers. In all cases, we find that in the early stages of the drying process $\dot{\epsilon}$  is nearly constant, with a numerical value that strongly depends on $Pe$. In this regime, the microscopic relaxation time of the NPs smoothly increases over several decades, reflecting the overall increase of their volume fraction in the shell. As drying proceeds, however, we observe two distinct regimes, depending on $Pe$. At relatively small $Pe \le 10$, $\dot{\epsilon}$ remains constant for the entire duration of the experiment, indicating that the NP mobility remains large enough to allow the drop to follow the contraction rate imposed by the drying conditions. In this regime, $\tau_D^{\rm{edge}}$ increases steadily as the overall NP concentration increases. By contrast, at larger $Pe$, the regime of constant $\dot{\epsilon}$ abruptly changes to a different regime characterized by a continuous decrease of $\dot{\epsilon}$ (solid black line in Fig.~\ref{fig:taudvsstarinrate}), throughout which the drop retains its nearly spherical shape. This regime lasts until the drop cannot shrink anymore in a uniform manner and the stress imposed by the solvent evaporation is relaxed through the shape instability (see Fig.~\ref{fig:imaging_top}, $t=1430$ s, and Fig.~\ref{fig:shell}a ii). The shrinking rate therefore displays a cross-over from an evaporation-limited regime at short times to a shell relaxation-limited regime at longer times.

Remarkably, we find that in the relaxation-limited regime and before the onset of the instability the microscopic relaxation time of the nanoparticles in the shell is very well described by a simple inverse proportionality law, $\tau_D^{\rm{edge}} = \Lambda \dot{\epsilon}^{-1}$, with a prefactor $\Lambda = (2.5 \pm 0.1)\times 10^{-4}$ independent of $Pe$. Because dynamic light scattering is sensitive to the relative displacement of scatterers, a possible origin of the $\tau_D^{\rm{edge}} \sim \dot{\epsilon}^{-1}$ scaling could be the difference in the displacement velocity of neighbouring portions of the shell associated to the same speckle, due to the radial dependence of the amplitude of the strain field. As detailed in~\cite{SM}, assuming that the local strain field is purely affine, the expected relaxation time would be $\tau_D^{\rm{edge}} \approx (2w q_y \dot{\epsilon})^{-1}$, implying $(2w q_y)^{-1} \equiv \Lambda_{\rm{aff}} = 7.5\times 10^{-4}$, with $w$ and $q_y$ the waist of the illuminating laser beam and the $y$ component of the scattering vector, respectively. This value is about three times larger than the experimental $\Lambda$, indicating that the microscopic dynamics are in fact dominated by non-affine displacements that grow proportionally to and add up to the underlying homogeneous strain field associated with the macroscopic contraction. These non-affine displacements speed up the relaxation dynamics, thereby allowing for efficiently reconfiguring the local structure in the shell, as required for the drop to keep reducing its radius.

The inverse proportionality relationship between $\tau_D^{\rm{edge}}$ and $\dot{\epsilon}$, together with values of $\phi_{\rm{shell}}$ above the glass transition volume fraction but below random close packing (see Fig.~\ref{fig:dynamics}c) and the lack of aggregation (see Figs.~\ref{fig:imaging_top} and~\ref{fig:dynamics}b), strongly supports the notion that the shell is a colloidal glass. Indeed, a similar scaling of the microscopic dynamics with the macroscopic deformation rate has been observed in other glassy and jammed materials~\cite{lee2009direct, brambilla_highly_2011,besseling2007three,varnik2006structural,laurati2012transient} submitted to an external drive. In all cases, the microscopic dynamics are dictated by the imposed deformation rate when the spontaneous dynamics are too slow to allow for structural relaxation on the time scale associated to the external drive.

\section{Discussion and Conclusions}
Our experiments on drying drops of colloidal suspensions unveil a richer scenario as compared to that currently accepted, built on experiments at very high $Pe \ge 10^{4}$ and based only on the notion of permanent aggregation of the NPs. Our findings highlight the crucial role of the shell visco-elasticity through its ability (or not) to relax the stress imposed by the macroscopic shrinking rate on the time scale dictated by the drying conditions. This ability is expected to depend on the drying rate, and thus on the Péclet number, but also on the volume fraction in the shell, which in turn depends on $Pe$. 

 \begin{figure}
\includegraphics[width=1\linewidth]{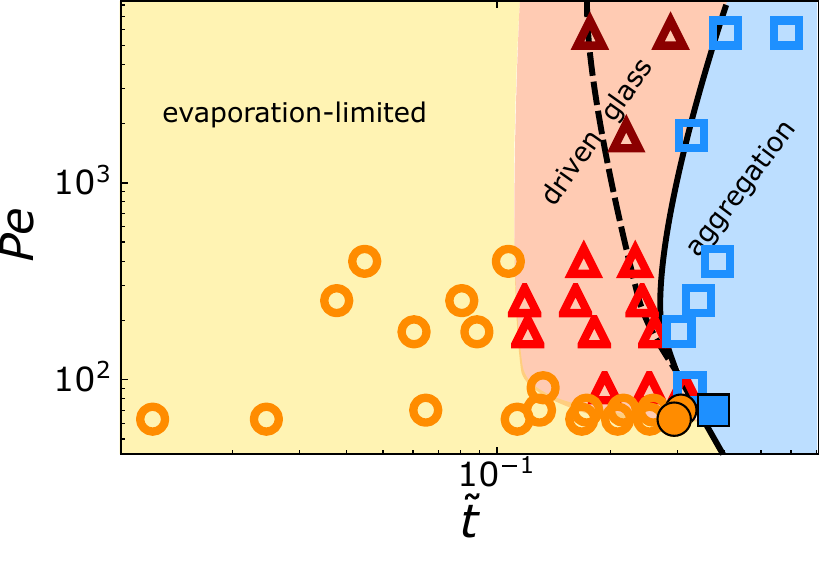}
\caption{\label{fig:phasediagram} \textbf{Unifying state diagram for the evolution of drying drops of colloidal suspensions {\color{myc} {sitting on a hydrophobic surface}}.}
The different regimes through which a drop of colloidal suspension transits while drying are depicted in the ($Pe$, $\tilde{t}$) plane. The dashed and solid lines indicate the onset of the shape instability and fracture instabilities, respectively. Yellow background and circle symbols: region where the drop shrinkage is ruled by the evaporation rate imposed by the environment relative humidity. Pink background and triangle symbols: driven glass region, where the drop evaporation is limited by the glassy dynamics of the shell. Blue background and square symbols: region where the particles in the shell are permanently aggregated. Open symbols indicate samples with a shell, full symbols at low $Pe$ designate samples with uniform particle distribution within the drop. The three topmost darker triangles correspond to the onset of shape instability for a sample that evaporated too fast for the microscopic dynamics to be measurable.}
\end{figure}


We propose an unifying scenario that rationalizes in a single state diagram in the ($Pe$, $\tilde{t})$ plane the results of all our experiments, as shown in Fig.~\ref{fig:phasediagram}. At relatively low $Pe \le 70$, {\textcolor{myc} {a shell is formed since the very early stage of drying ($\tilde t = 0.169$ for $Pe = 63$ and $\tilde t = 0.129$ for $Pe = 70$)}} (open circles), but the evaporation rate is slow enough for the NPs to recover a uniform distribution within the drop at larger $\tilde{t}$ (filled symbols). In this regime, the contraction rate $\dot{\epsilon}$ is constant and imposed by the relative humidity of the environment (see semi-filled symbols in Fig.~\ref{fig:taudvsstarinrate}). Eventually, a fracture instability occurs, when the suspension is too dense to allow for further drop contraction. The fracture instability occurs when the system crosses the boundary between a dispersed and an aggregated state, respectively; this boundary is indicated in Fig.~\ref{fig:phasediagram} by the solid line to the left of the aggregation region. 

At larger $Pe > 70$, the drops initially follow the same evaporation-limited evolution as for lower $Pe$. However, they subsequently enter a driven-glass regime, where the drop contraction slows down and $\tau_D^{\rm{shell}}$, the microscopic relaxation time of the shell, and $\dot{\epsilon}$ are mutually slaved, as shown by the solid line in Fig.~\ref{fig:taudvsstarinrate}. In this regime, we find that $\Lambda = \tau_D\dot{\epsilon}$ is essentially independent of $Pe$ and $\tilde{t}$, although both parameters have an impact on $\phi_{\mathrm{shell}}$. This is consistent with the glassy nature of the shell: once the rearrangement dynamics driven by the macroscopic shrinkage become faster than the spontaneous ones, the structure and quiescent dynamics in the shell become irrelevant, the driving rate $\dot{\epsilon}$ being the only relevant parameter. As the shell densifies, the drops undergo a first shape instability, by forming an invagination that allows for further water evaporation. This first instability is indicated by the dashed line in the state diagram of Fig.~\ref{fig:phasediagram} and is well distinct from the fracture instability occurring at later $\tilde{t}$ (solid line). Indeed, we find that no particle aggregation occurs in the driven glass regime and across the shape instability, while in all fractured samples the NPs are permanently aggregated. {\textcolor{myc} {Intuitively, one would expect that a prerequisite for cracking is particle aggregation and our work clearly indicates that this is indeed the case for our system.}

{\textcolor{myc} {Figures~\ref{fig:taudvsstarinrate}} and~\ref{fig:phasediagram} show that the driven glass regime does not extend to arbitrarily low $Pe$, since it is not observed for $Pe \leq 70$. {\textcolor{myc} {Hence, despite similar features of the shell, compare e.g.  the shell thickness (Fig.~\ref{fig:shell}d) and volume fraction (Fig.~\ref{fig:dynamics}c), the drops dried at $Pe=70$ and $Pe=91$ behave differently, because the shells are compressed at different rates.}} At first sight, this is quite puzzling, since one may think that, whatever the drying rate and hence $Pe$, as $\phi$ grows beyond $\phi_g$ the microscopic dynamics would always become too slow for the drop to maintain the initial $\dot{\epsilon}$. Figure~\ref{fig:taudvsstarinrate} shows that drops that avoid the driven glass regime contract at a rate lower than a threshold $\dot{\epsilon}_{th} \approx 1.5 \times 10^{-5}~\mathrm{s}^{-1}$, suggesting that our NP suspensions can accommodate ultraslow structural relaxations on time scales of the order of $1/\dot{\epsilon}_{th} \approx 6.7 \times 10^{4}~\mathrm{s}$. This is indeed confirmed by the measurements on quiescent suspensions of Ref.~\cite{philippe2018glass}: at large $\phi \gtrsim \phi_g$, the structural relaxation time exhibits a very weak $\phi$ dependence and plateaus to $\tau_D \approx 2 \times 10^4~\mathrm{s}$, a value of the same order of magnitude of $1/\dot{\epsilon}_{th}$. This peculiar dynamic regime has been observed for a variety of systems interacting through a repulsive interparticle  potential softer than the infinitely steep hard sphere potential~\cite{philippe2018glass,srivastava2013structure,li2017long}, as is the case of the screened Yukawa potential of our charge-stabilized NPs.

Note that the shape of the various regions in the state diagram of Fig.~\ref{fig:phasediagram} changes if one uses the physical time $t$, rather than the reduced time $\tilde{t}$, as the abscissa, see Fig.~{\color{myc} {14}} in~\cite{SM}. In particular, the driven glass region becomes increasingly narrower as $Pe$ grows, since the aggregation stage is reached very rapidly. The duration of the glass region lasts about $400$ s for $Pe=91$ and decreases down to $50$ s for $Pe=5754$, see Fig.~{\color{myc} {15}} in~\cite{SM}. This is likely to explain why the driven glass regime and the occurrence of two distinct instability could not be resolved in previous works on very fast-evaporating drops~\cite{tsapis2005onset,lintingre2015controlling,lintingre2016control,lyu2019final}. We expect the exact location of the boundaries between the different regions to depend also on other control parameters that are non explicitly included in the state diagram of Fig.~\ref{fig:phasediagram}, such as the nature of the substrate, the interparticle potential, and the initial volume fraction. We confirm this hypothesis by performing additional experiments, changing the contact angle between the drop and the surface, the NP volume initial fraction and the range of the repulsive potential between nanoparticles, see~\cite{SM}. We find that varying these parameters does indeed shift the boundaries of the state diagram and modifies the detailed morphology of the shape instability, an effect previously reported for the shape instability observed in 2D experiments~\cite{basu_towards_2016,pauchard_invagination_2004}. Crucially, however, the physical picture of a double state transition is not altered, strongly suggesting that the scenario proposed here is robust.

In conclusion, we have shown that drops of colloidal suspensions drying at intermediate {\color{myc2} {to large}} $Pe$ exhibit a richer-than-expected behavior, generally going through two distinct shape instabilities, which we rationalize in the framework of a unified state diagram. These instabilities are similar in that they both stem from the formation of a dense shell, which eventually becomes too rigid to accommodate the drop retraction imposed by drying. However, rigidity originates from two totally distinct mechanisms: a repulsive glass transition first, and the irreversible aggregation of NPs at larger times. 
{\color{myc} Of note our findings might be reminiscent of the two-state drying identified in directional drying of thin film, with a first state corresponding to a reversible colloidal crystallization and the second state corresponding to irreversible colloidal aggregation \cite{goehring_solidification_2010}}.

Besides providing a deeper understanding of the drying mechanisms of colloidal drops at the fundamental level, we expect our work to have implications in applications involving relatively slow drying, e.g. in the synthesis of functional supraparticles~\cite{liu2019tuning,kuncicky_surface-guided_2008,marin_building_2012,sperling2014controlling,sekido2017controlling,kong2022virus}, or the drying of drops containing biological entities, e.g. viruses or bacteria~\cite{huynh2022evidence,xie_bacterial_2006,grinberg2019bacterial}. For the former, the finding that shape instabilities and shell formation occur at a stage where the NPs may be fully re-dispersed potentially paves the way to the synthesis of supraparticles with new morphologies and properties, e.g. by varying the evaporation rate during the processing. For the latter, the formation of a reversible crust that protects against further evaporation while allowing for re-dispersion is likely to may play an important role in the survival, e.g., of cells embedded in drops undergoing cyclic variations between wet and dry conditions~\cite{grinberg2019bacterial,hoefman2012survival,wesche2009stress}.

\section*{Author Contributions}
We strongly encourage authors to include author contributions and recommend using \href{https://casrai.org/credit/}{CRediT} for standardised contribution descriptions. Please refer to our general \href{https://www.rsc.org/journals-books-databases/journal-authors-reviewers/author-responsibilities/}{author guidelines} for more information about authorship.

\section*{Conflicts of interest}
There are no conflicts to declare.

\section*{Acknowledgements}
We acknowledge financial support from the French Agence Nationale de la Recherche (ANR) (Grant No. ANR-19-CE06-0030-02, BOGUS). LC gratefully acknowledges support from the Institut Universitaire de France. We thank L. Pauchard and F. Giorgiutti-Dauphiné for discussion. 



\bibliography{rsc} 
\bibliographystyle{rsc} 

\end{document}